\documentclass[prb,preprint,amsmath,amssymb]{revtex4}

\usepackage{graphicx}
\usepackage{natbib}
\usepackage{amsmath}
\usepackage{mathrsfs}

\usepackage{subfigure}
\usepackage{epsfig} 
\usepackage{epstopdf}
\usepackage{float} 
\usepackage{tikz}
\usetikzlibrary{shapes.arrows}
\usepackage[percent]{overpic}
\usepackage{comment}
\usepackage{pstricks}
\usepackage{setspace}

\begin{document}

\newcommand{\Hm}{{\cal H}}
\newcommand{\Tr}{\hbox{Tr}}
\newcommand{\Rneq}{\rho_{\hbox{\scriptsize neq}}}
\newcommand{\Pneq}{\Psi_{\hbox{\scriptsize neq}}}
\newcommand{\Peq}{\Psi_{\hbox{\scriptsize eq}}}
\newcommand{\Ahat}{\hat{A}}
\newcommand{\Vhat}{\hat{V}}

\title{Regression relation for pure quantum states and its implications for efficient computing}

\author{Tarek A. Elsayed}
\email{Tarek.Ahmed.Elsayed@gmail.com}
\address{Institute for Theoretical Physics, University of Heidelberg, Philosophenweg 19, 69120 Heidelberg, Germany}
\address{Electronics Research Institute, Dokki, Giza 12622, Egypt}
\author{Boris V. Fine}
\thanks{Corresponding author}
\email{B.Fine@thphys.uni-heidelberg.de}
\address{Institute for Theoretical Physics, University of Heidelberg, Philosophenweg 19, 69120 Heidelberg, Germany}

\begin{abstract}
We obtain a modified version of the Onsager regression relation for the expectation values of quantum-mechanical operators in pure quantum states of isolated many-body quantum systems. We use the insights gained from this relation to show that high-temperature time correlation functions in many-body quantum systems can be controllably computed without complete diagonalization of the Hamiltonians, using instead the direct integration of the Schroedinger equation for randomly sampled pure states.
This method is also applicable to quantum quenches and other situations describable by time-dependent many-body Hamiltonians. The method implies exponential reduction of the computer memory requirement in comparison with the complete diagonalization. 
We illustrate the method by numerically computing infinite-temperature correlation functions for translationally invariant Heisenberg chains of up to 29 spins 1/2. Thereby, we also test the spin diffusion hypothesis and find it in a satisfactory agreement with the numerical results. Both the derivation of the modified regression relation and the justification of the computational method are based on the notion of quantum typicality.

\end{abstract}


\date{22 August, 2012}

\keywords{correlation function, onsager regression hypothesis,strongly correlated system}
\maketitle

In 1931, Onsager came up with the profound insight that ``the average regression of fluctuations will obey the same laws as the corresponding macroscopic irreversible process" \cite{Onsager-31}. This statement known as the ``Onsager regression hypothesis'' (ORH) became the cornerstone of the linear response theory.  From today's perspective, the ORH is equivalent\cite{Ford-96,Lax-00} to the high-temperature limit of the fluctuation-dissipation theorem\cite{Callen-51}.  In this paper, we show that a modified version of ORH holds for the expectation values of quantum-mechanical operators, when a many-body system is in a pure quantum state. We also present an efficient method for computing high-temperature linear response characteristics of many-particle quantum systems using the time evolution of a single pure state.

There exists a class of nonperturbative problems, such as nuclear spin-spin relaxation in solids\cite{Abragam-61}, where the relaxation or correlation functions in translationally-invariant systems need to be computed at high temperatures. Despite the progress in the approximate methods, e.g. \cite{Fine-97,Zhang-07}, and numerical techniques\cite{Verstraete-04,White-04,Schollwock-05,Vidal-08}, the above kind of problems generally resist {\it controllable} solutions, leaving the complete diagonalization of quantum Hamiltonians as the only way to obtain controllable results. The sizes of the systems treatable by complete diagonalization are severely limited by the computer memory requirement that scales as $N^2$, where $N$ is the number of quantum states in the system. The memory requirement for the controllable-accuracy algorithm proposed in this work  scales at most a $N (\log N)^2$.

In recent years,  it was realized that, given the exponentially large number $N$ of quantum states in a many-particle system, many observable properties of such a system can be obtained by sampling one suitably chosen pure quantum state, or a wave function --- the so-called ``quantum typicality''\cite{Gemmer-04,Goldstein-06,Popescu-06,Reimann-07}. In particular, Refs.\cite{Alvarez-08,Bartsch-09,Bartsch-11} applied the notion of quantum typicality to the relaxation and fluctuation phenomena, but on the numerical side these investigations dealt so far only with the systems that were sufficiently small to perform the complete diagonalization of the Hamiltonians. 

In this paper, we report a conceptual and a computational results, which are both connected to the notion of typicality but, otherwise, only indirectly connected to each other. The conceptual result is that the expectation values of quantum-mechanical operators in a {\it pure} quantum state obey the usual regression relation but with the amplitude of fluctuations exponentially reduced in comparison with the classical case (see Eq.(\ref{regres}) below). The computational result is that the high-temperature time correlation functions can be controllably computed on the basis of  Eq.(\ref{fluct1}) without complete diagonalization of the Hamiltonians, using instead the direct integration of the Schroedinger equation for randomly sampled pure states. As an example, we obtain  infinite-temperature correlation functions for translationally invariant Heisenberg chains of up to 29 spins 1/2, thereby also testing the spin diffusion hypothesys. To the best of our knowledge, none of the complete diagonalization studies of the Heisenberg spin-1/2 chains conducted so far has reached the above size. 
We note here that pure quantum states were used in Refs.\cite{Long-03,White-09} in the context of other numerical methods.

Below, in order to be specific, we consider a lattice of $N_s$ interacting spins 1/2 with the total number of quantum states $N = 2^{N_s} \gg 1$ and the Hamiltonian ${\cal H}$. We adopt the following conventions: (i)  Analytical formulas are presented only in the leading order in $1/N$. (ii) Wave functions without time argument and operators with time argument imply the Heisenberg representation of quantum mechanics. The opposite implies the Schr\"{o}dinger representation. (iii) $ \hbar=1 $.

We now focus on some observable quantity, e.g. total spin polarization, characterized by operator $\Ahat \equiv A_{mn}$, which has zero average value at the infinite-temperature equilibrium, i.e. $\Tr \{ \Ahat \} = 0$.
The Onsager regression relation for this quantity near the infinite temperature equilibrium has the following form:
\begin{equation}
\Tr \left\{ \Ahat(t) \Rneq \right\} = {\alpha \over N} \Tr \left\{ \Ahat(t) \Ahat(0) \right\} .
\label{Onsager}
\end{equation}
where $\Ahat(t) = e^{i \Hm t} \Ahat e^{-i \Hm t}$, and $\Rneq = {1 \over N} \exp (  \alpha \Ahat )$ with $\alpha$ being a small constant. The right-hand side (RHS) of Eq.(\ref{Onsager}) represents the equilibrium time correlation function of $\Ahat$, while the left-hand side (LHS) is the relaxation function of $\Ahat$ corresponding to the initial nonequilibrium density matrix $\Rneq$. 

Quantum typicality investigation of Ref.\cite{Bartsch-09} implied that
\begin{equation}
\langle \Pneq | \Ahat(t) |  \Pneq \rangle = \Tr \left\{ \Ahat(t) \Rneq \right\}
\left[1 + O\left( {1 \over \alpha \sqrt{N}} \right) \right] ,
\label{relax}
\end{equation}
where $|\Pneq \rangle$ is a wave function that ``samples'' $\Rneq$.  

Now we obtain a complementary relation on the fluctuation side. It involves the wave function $| \Peq \rangle$ representing the infinite temperature equilibrium and defined as a random vector in the Hilbert space of the system. $| \Peq \rangle$ can be generated in {\it any} orthonormal basis $\{ |\phi_k \rangle \}$ as follows:
\begin{equation}
 |\Peq \rangle = \sum_{k=1}^N a_k | \phi_k \rangle ,
\label{Psi-eq}
\end{equation}
where  $a_k$ are the quantum amplitudes, whose absolute values are selected from the probability distribution\cite{Fine-09,Normalization}
\begin{equation}
P(|a_k|^2) = N \exp (- N |a_k|^2)
\label{P}
\end{equation}
and the phases are chosen randomly in the interval $[0, 2\pi)$.
In the following, we use bar above an expression to indicate the Hilbert-space average over all possible choices on $|\Peq \rangle$.

Now we consider the correlation function for the time series of the expectation value $\langle \Peq | \Ahat (t')| \Peq \rangle$ in the time interval $[-T, T+t]$:
\begin{equation}
C(t, T) \equiv {1 \over 2 T} \int_{-T}^{T} dt' \langle \Peq | \Ahat (t+t')| \Peq \rangle \langle \Peq | \Ahat(t') |  \Peq \rangle .
\label{C}
\end{equation}
In \cite{Supplement} we derive the following relation:
\begin{equation}
C(t, T) = {1 \over N^2}  \Tr \left\{ \Ahat(t) \Ahat(0) \right\} +  \Delta_1 ,
\label{fluct2}
\end{equation}
where   
\begin{widetext}
\begin{equation}
\overline{\Delta_1^2} \approx
{1 \over 2\sqrt{2} \ T N^4} \int_{-T\sqrt{2}}^{T\sqrt{2}} dt_2 
\left( 
\left[ \Tr \left\{ \Ahat(t_2) \Ahat(0) \right\} \right]^2
+ \Tr \left\{ \Ahat (t- t_2) \Ahat(0) \right\} \Tr \left\{ \Ahat (t+ t_2) \Ahat(0) \right\}
\right).
\label{Otau}
\end{equation}
\end{widetext}
For large enough $T$, the correction $\Delta_1$ in  Eq.(\ref{fluct2}) is much smaller than the principal term as long as $\Tr \left\{ \Ahat(t) \Ahat(0) \right\} \xrightarrow{\scriptscriptstyle t\to\infty} 0$. In particular, if $\Tr \left\{ \Ahat(t) \Ahat(0) \right\}$ decays at large $t$ faster than $|t|^{-0.5}$ then, for large enough $T$, the integral in Eq.(\ref{Otau}) becomes independent of $T$, and, as a result, $\Delta_1 = O(\sqrt{\tau/T}) \Tr \left\{ \Ahat^2 \right\}/N^2$, where  $\tau$ is the characteristic timescale for the decay of the expression under the integral. If $\Tr \left\{ \Ahat(t) \Ahat(0) \right\}$  decays asymptotically as $|t|^{-\nu}$ with $0 < \nu < 0.5$, then, according to Eq.(\ref{Otau}),  $\Delta_1$ still remains small, but its prefactor scales as $O(T^{-\nu})$.  

Eq.(\ref{fluct2}) together with Eqs.(\ref{Onsager},\ref{relax}) implies the modified version of ORH for the expectation value of the operator $\Ahat$ in a pure quantum state:
\begin{equation}
\lim_{N \to \infty} \langle \Pneq | \Ahat(t) |  \Pneq \rangle = \alpha \ \lim_{T \to \infty, N \to \infty} N C(t,T),
\label{regres}
\end{equation}
where, in the RHS, the limit $N \to \infty$ is taken first, which in practical terms means that $T$ should be much smaller than the inverse spacing of the energy levels in the system as the above limits are taken\cite{Supplement}.

From the viewpoint of practical computing, the implications of the above findings are two-fold: 
(i) As already implicit in Eq.(\ref{relax}), and explicit in Eqs.(\ref{fluct2},\ref{regres}), a single realization of $\langle \Pneq | \Ahat(t) |  \Pneq \rangle$ is exponentially more accurate in approximating $\Tr \left\{ \Ahat(t) \Ahat(0) \right\}$ than the corresponding single classical relaxation process in approximating classical correlation function\cite{Supplement}. $\langle \Pneq | \Ahat(t) |  \Pneq \rangle$ decays into the equilibrium statistical noise $\langle \Peq | \Ahat(t) |  \Peq \rangle$, which, according to Eq.(\ref{fluct2}) has root-mean-squared (rms) amplitude $\sqrt{C(0,T) } \approx \sqrt{\Tr \{ \Ahat^2 \}}/N$, which is by factor of $\sqrt{N}$ smaller than the rms amplitude $\sqrt{\Tr \{ \Ahat^2 \}/N}$ expected for the classical noise or the noise of continuously monitored macroscopic quantum observable at infinite temperature\cite{Bloch-46,Supplement}. This noise suppression is due to the fact that the time evolution of a single pure state contains  the superposition of $N$ independent ``noises'' associated with each of the basis states\cite{Alvarez-08}.  The statistical noise of $\langle \Pneq | \Ahat(t) |  \Pneq \rangle$ can be suppressed further by averaging over many pure-state evolutions. (ii) In principle, as we show below, the direct evaluation of $C(t,T)$  can also be used to obtain $\Tr \left\{ \Ahat(t) \Ahat(0) \right\}$, but this procedure does not take advantage of the above-mentioned quantum parallelism and hence is less efficient.

Although the evaluation of $\langle \Pneq | \Ahat (t)|\Pneq \rangle$ is a very efficient method to obtain $\Tr \left\{ \Ahat(t) \Ahat(0) \right\}$, an even more efficient method is to use  typicality to sample this trace directly on the basis of the following relation derived in \cite{Supplement}:
\begin{equation}
\langle \Peq | \Ahat (t) \Ahat(0) |  \Peq \rangle = {1 \over N}  \Tr \left\{ \Ahat(t) \Ahat(0) \right\}  + \Delta_2   ,
\label{fluct1}
\end{equation}
where
\begin{equation}
\overline{\Delta_2^2} = {1 \over N^2} \Tr \left\{ \Ahat(t) \Ahat(0) \Ahat(t) \Ahat(0) \right\} .
\label{epsilon}
\end{equation}
That the second term in the RHS of Eq.(\ref{fluct1}) is much smaller than the first one can be shown by estimating their ratio at $t=0$ as
${\sqrt{\Tr \left\{ \Ahat^4 \right\} } \over \Tr \left\{ \Ahat^2 \right\} } \sim {1 \over \sqrt{N}} $. The statistical accuracy of computing $\Tr \left\{ \Ahat(t) \Ahat(0) \right\}$ with the help of Eq.(\ref{fluct1}) is thus better by factor $1/\alpha$ in comparison with Eq.(\ref{relax}).

\begin{figure} \setlength{\unitlength}{0.1cm}
\begin{picture}(88 , 60 )
{
\put(-3, 0){ \epsfig{file=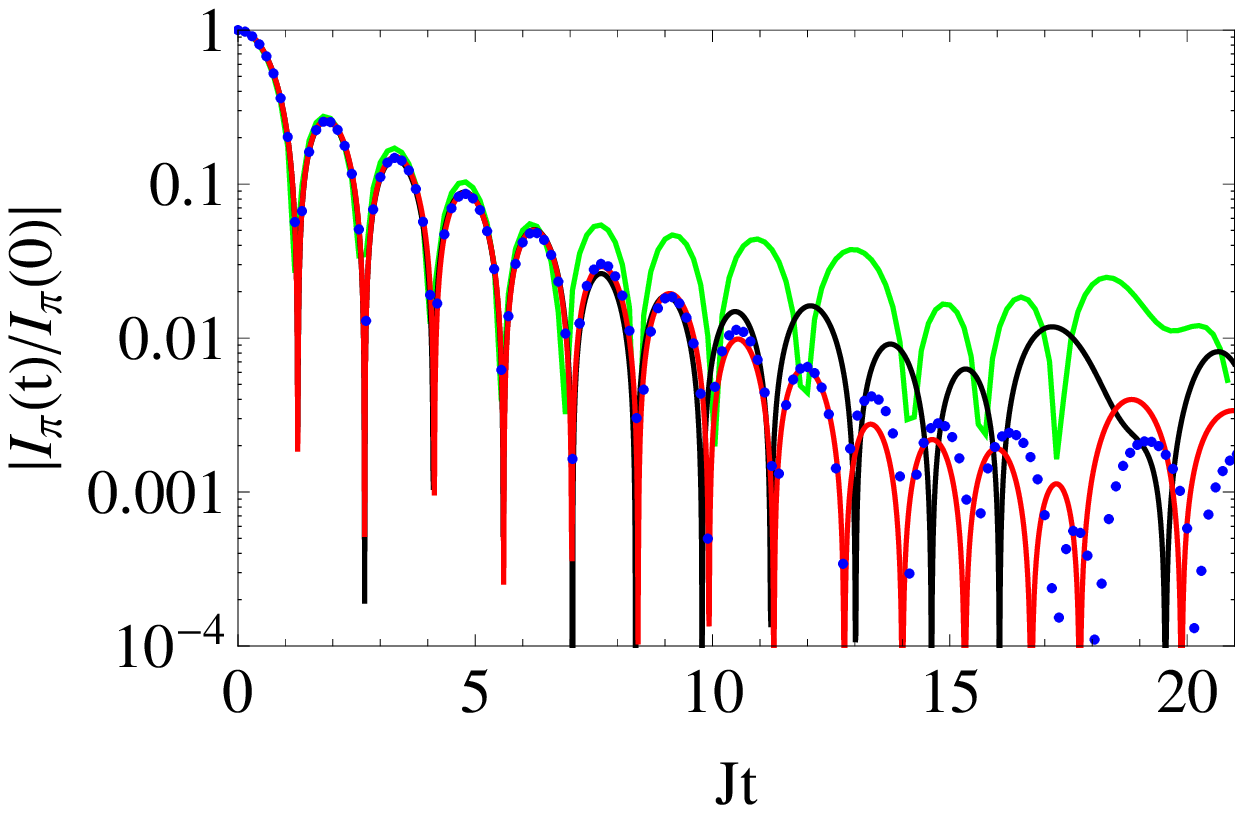,width=8.8cm } }
\put(56, 40){ \epsfig{file=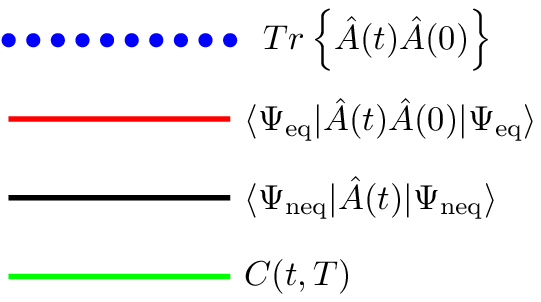,width=2.7cm } }
}
\end{picture}
\caption{ \label{fig-test} Intermediate dynamic structure factor $I_{\pi}(t)$ of the Heisenberg chain of 20 spins 1/2. The calculation based on the exact trace formula (\ref{I}) is compared with the approximations given by Eqs.(\ref{relax},\ref{fluct2},\ref{fluct1}) as indicated in the legend. The initial agrement between the black and the green lines also demonstrates the validity of the regression relation (\ref{regres}). All calculations are based on the complete diagonalization of the Hamiltonian $\Hm$. Each of the three approximate calculations is done with a single pure state. In the case of Eq.(\ref{relax}), $\alpha = 0.083$ corresponding to the initial polarization equal to approximately 4 percent of the maximum polarization. In the case of Eq.(\ref{fluct2}), $T=4200/J$. As expected theoretically, the approximation based on Eq.(\ref{fluct1}) gives the most accurate agreement with the exact result. (Note the logarithmic vertical scale.) The accuracy of all three approximations can be improved by averaging over more pure states.  
}
\end{figure}

In Fig.\ref{fig-test} we demonstrate the relationships (\ref{relax}, \ref{fluct2}, \ref{fluct1}) by computing the intermediate dynamic structure factor $I_{\pi}(t)$ for the Heisenberg chain of 20 spins 1/2 using complete diagonalization. Thereby we also demonstrate the regression relation (\ref{regres}). The Hamiltonian of this chain is  $\Hm = J \sum_{i} \mathbf{S_i}\cdot \mathbf{S_{i+1}} $ with periodic boundary condition. Here  $J$ is the coupling constant, and $\mathbf{S_i}$ is the spin operator on the $i$th chain site. Such a chain admits periodic spin modulations with wave numbers $q = 2 \pi n/N_s$, where $n$ is an integer number taking values $0, 1, ..., N_s-1$. For a given wave number $q$, the intermediate dynamic structure factor is defined as
\begin{equation}
I_q(t) \cong  \Tr \left\{\Ahat_{ \{q\} }(t) \  \Ahat_{ \{q\} }(0)  \right\},  
\label{I}
\end{equation} 
where $\Ahat_{ \{ q \} } = \sum_m \cos(q m) S_m^x$.

Now, we proceed with showing that, for the spin systems too large to be treated by complete diagonalization, it is still possible to controllably compute infinite temperature correlation functions  by evaluating the LHS of Eq.(\ref{fluct1}) with the help of the direct integration of the Schr\"{o}dinger equation.

We compute the time evolution of pure states on the basis of the time-discretized version of the Schr\"{o}dinger equation. We use the fourth-order Runge-Kutta routine based on the following equation:
\begin{equation}
|\Psi(t + \Delta t)  \rangle = |\Psi(t) \rangle + |v_1\rangle + |v_2\rangle + |v_3\rangle + |v_4\rangle,
\label{discrete}
\end{equation}
where $\Delta t$ is the discretization time step, and $|v_1\rangle,|v_2\rangle, |v_3\rangle, |v_4\rangle$ are unnormalized Hilbert-space vectors computed as follows:
$|v_1\rangle = -i \Hm |\Psi(t)\rangle \Delta t$, $|v_2\rangle = -{1 \over 2} i \Hm |v_1\rangle \Delta t$, $|v_3\rangle = -{1 \over 3} i \Hm |v_2\rangle \Delta t$, and $|v_4\rangle = -{1 \over 4} i \Hm |v_3\rangle \Delta t$. Given the linearity of the Schr\"odinger equation, Eq.(\ref{discrete}) is  equivalent to the simple 4th order power-series expansion of the time-evolution operator at each discretization time step. We used $\Delta t = 0.01/J$ 

The above routine requires only storing in the memory the vectors $|\Psi\rangle$ and $|v_i\rangle$ and the non-zero elements of the Hamiltonian $\Hm$. Although the Hamiltonian is an $N \times N$ matrix, it is very sparse for many-particle systems with only two-particle interactions when represented in a ``local'' basis, where each basis function is factorizable in terms of the wave functions of individual particles. For $N_s$ spins 1/2, a possible local basis is the one where the $z$-projections of all spins are quantized. In this basis, the number of the nonzero entries of the Hamiltonian matrix is of the order of $N \times N_s^2$ for the systems with long-range interactions, or $N \times N_s$ for the short range interactions. Thus the overall memory required for the direct propagation of the Schroedinger equation scales at most as $N (\log N)^2$, i.e. it is exponentially smaller than the memory required for the complete diagonalization, which scales as $N^2$. One can take advantage of this memory reduction only when the operator of interest, $\Ahat$, is also sparse in the local basis, but this is normally the case in physical contexts. In fact, in many cases, including the calculations of $I_q(t)$, it is possible simply to use the eigenbasis of $\Ahat$ as the local basis.

\begin{figure} \setlength{\unitlength}{0.1cm}

\begin{picture}(88 , 25 ) 
{
\put(0, 0){  \epsfig{file=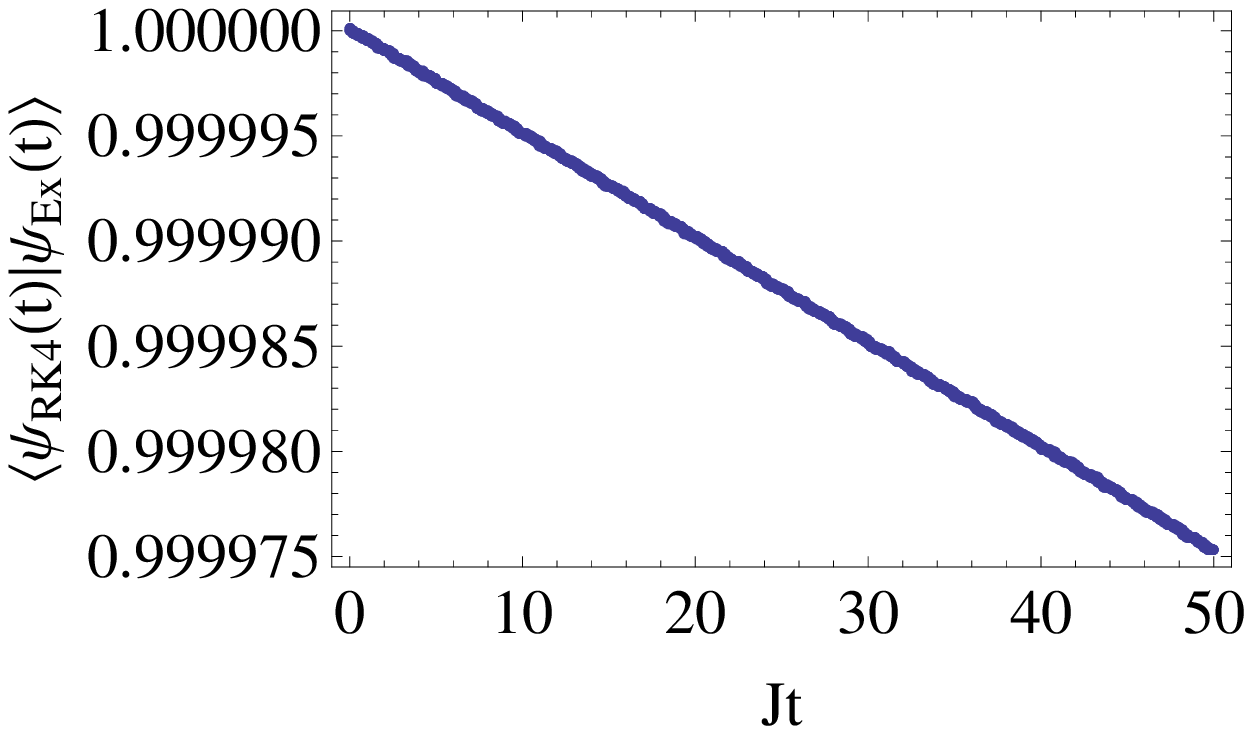,width=4cm } }
\put(44, 0){  \epsfig{file=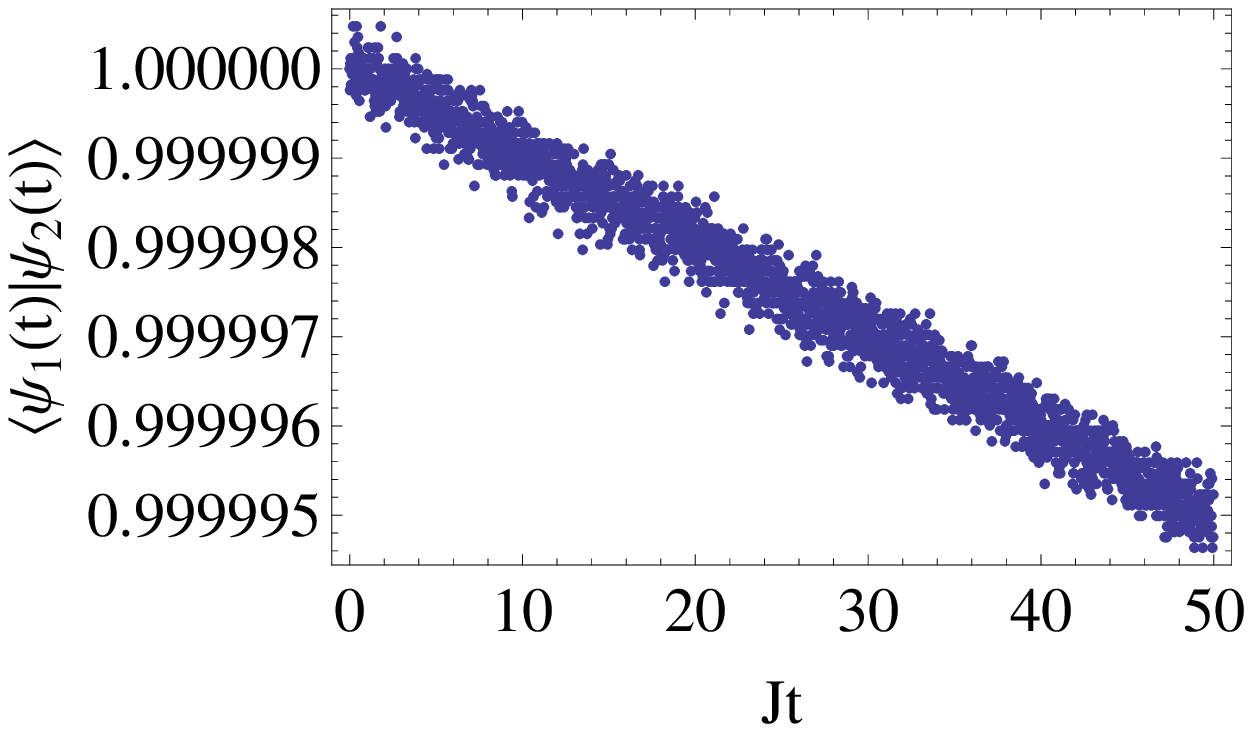,width=4cm } }
\put(35,19){{(a)}}
\put(79,19){{(b)}}
}
\end{picture} 
\caption{ \label{fig-test2}  Tests of numerical accuracy of the direct integration of the Schr\"odinger equation. (a) Overlap between two initially identical wave functions for the Heisenberg chain of 20 spins 1/2, $| \Psi_{\hbox{\scriptsize Ex}}(t) \rangle$ and $| \Psi_{\hbox{\scriptsize RK4}}(t) \rangle$,  computed using, respectively, the complete diagonalization and the direct integration. (b) Overlap between two initially identical wave functions for the Heisenberg chain of 29 spins 1/2, $| \Psi_1(t) \rangle$ and $| \Psi_2(t) \rangle$, both computed using the direct integration method with two respective discretization time steps $\Delta t_1 = 0.01/J$ and $\Delta t_2 = 0.02/J$. The noisiness of the line originates from the accumulated machine rounding errors.
}
\end{figure}

We verify the accuracy of the direct integration method in two ways. For small spin clusters, we compare the wave functions obtained by propagating the same initial state using either complete diagonalization or the direct integration method. As shown in Fig.~\ref{fig-test2}(a), the overlap between the two wave functions remains extremely close to 1 over the time interval required to compute $I_{\pi}(t)$ for 20 spins 1/2 in Fig.~\ref{fig-test}. For larger systems, we compare two wave functions $\vert\Psi_1 (t)\rangle$ and $\vert\Psi_2 (t)\rangle$ obtained by propagating the same initial wave function using the direct integration method with two different discretization time steps $\Delta t_1$ and $\Delta t_2$ such that $\Delta t_2 = 2 \Delta t_1$.  We then verify that  that their overlap   $\left \langle \Psi_1 (t)|\Psi_2 (t) \right \rangle$ is sufficiently close to 1. An example of such a test for 29-spin Heisenberg chain is shown in Fig.~\ref{fig-test2}(b). 

We note here that the same direct integration algorithm can be used to compute the imaginary-time evolution associated with the expression $\exp(- \Hm \beta/2) |\Peq \rangle$, where $\beta$ is the inverse temperature, thereby generating equilibrium wave function corresponding to temperature $1/\beta$. This wave function can then be used to compute the linear response characteristics at temperature $1/\beta$.

Our numerical procedure for computing $\langle \Peq | \Ahat (t) \Ahat(0) |  \Peq \rangle$ is based on propagating two wave functions using the direct integration method. One of them is $|\Peq(t) \rangle = \exp(-i \Hm t) |\Peq(0) \rangle$, where $|\Peq(0) \rangle$ is given by Eq.(\ref{Psi-eq}). The other one is 
$|\Phi(t) \rangle = \exp(-i \Hm t) |\Phi(0) \rangle$, where $|\Phi(0) \rangle = \Ahat |\Peq(0) \rangle $ (i.e. $|\Phi(0) \rangle$ is unnormalized).  The quantity of interest $\langle \Peq | \Ahat (t) \Ahat(0) |  \Peq \rangle$ is then evaluated as $\langle \Peq(t) | \Ahat |\Phi(t) \rangle $.

Now we exemplify the direct integration method by computing the intermediate dynamic structure factors $I_q(t)$ with $q = 2\pi/N_s$ for Heisenberg chains of sizes up to $N_s = 29$. 
By doing this calculation, we also test the spin diffusion hypothesis, which stipulates that, for sufficiently small values of $q$,   $I_q^{\mu}(t)\approx e^{-D q^2 t}$, where $D$ is the diffusion coefficient. 

\begin{figure} \setlength{\unitlength}{0.1cm}
\begin{picture}(88 , 60 )
\put(-2,0){ \epsfig{file=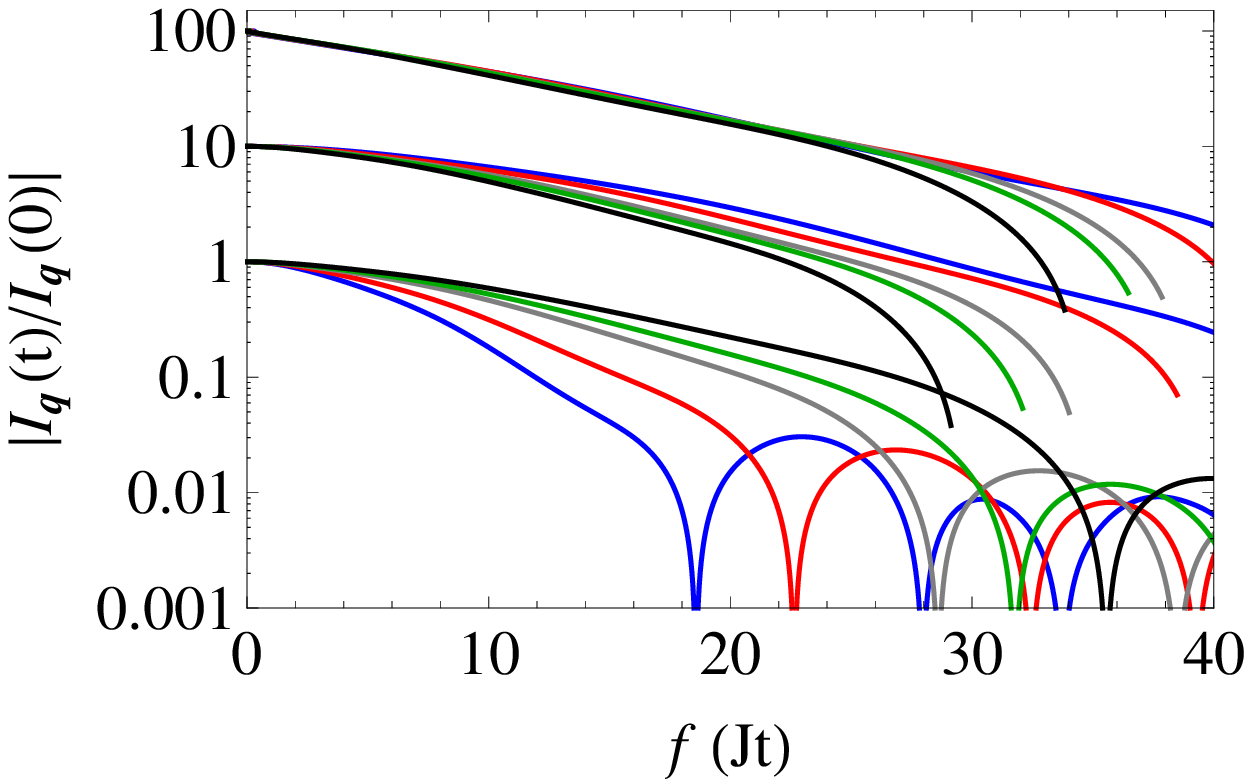,width=8.8cm } }
\put(17,13){ \epsfig{file=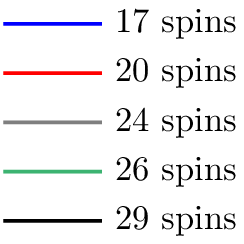,width=1.7cm } }
\put(31,36){{(a)}}
\put(31,45){{(b)}}
\put(31,51){{(c)}}
\end{picture}
\caption{ \label{fig-Iq} Intermediate dynamic structure factors $ I_{q}( t )$ for Heisenberg chains of $N_s$ spins 1/2 computed as the LHS of Eq.(\ref{fluct1}) by propagating a single pure state with the help of the direct integration of the Schr\"odinger equation. The values of $N_s$ are indicated in the plot legend. In each case, $q=2\pi/N_s$. The horizontal axis is: (a) $f(Jt) = Jt$, (b) $f(Jt) = \xi_1 \ q^2 t$, and (c) $f(Jt) = \xi_2 \ q^2\left( 1+0.1 \ \hbox{ln} q \right) t \hbox{ln} t $, where $\xi_1$ and $\xi_2$ are arbitrary scaling parameters. The vertical axes for (b) and (c) are displaced for better visibility. Plots (a) represent the original calculation results. Plots(b) test the scaling expected for spin diffusion. Plots (c) test the empirical scaling reported for classical spins in Ref.\cite{Bonfim-92}. The numerical accuracy test for the pure state evolution in the case of the 29-spin chain is given in Fig.~\ref{fig-test2}(b).}
\end{figure}

Our results presented as plots (a) in Fig.\ref{fig-Iq} indicate that, in every case, $I_q(t)$ shows the initial tendency to decay exponentially, but then the behavior universally starts exhibiting oscillations. Plots (b) in the same figure further indicate that the nearly exponential parts of $I_q(t)$ exhibit satisfactory $q^2$-scaling, while plots 2(c) show that  the scaling $q^2 (1 + 0.1 \ \hbox{ln} q) \hbox{ln} t $ reported in the numerical studies of classical spins\cite{Bonfim-92} works even better. 

To summarize, we obtained the modified Onsager regression relation (\ref{regres}) for  a pure quantum state. We also find that the direct computation of the LHS of Eq.(\ref{fluct1}) is the most efficient way to obtain equilibrium time correlation functions with controllable accuracy. 
We have directly tested only the high-temperature limit but the method itself can also be used at finite temperatures. We further note that the direct integration of the Schr\"odinger equation in combination with the random sampling of pure states can also be used for the efficient computing of quantum quenches and other situations describable by time-dependent many-body Hamiltonians.

B.F is grateful to P. Schmitteckert for discussions related to this paper. B.F. also acknowledges the hospitality of the Kavli Institute for Theoretical Physics at the University of California, Santa-Barbara, where a part of this paper was written and supported by the National Science foundation under Grant No. NSF PHY11-25915. The numerical part of this work was performed at the bwGRiD computing cluster at the University of Heidelberg. 


\begin{thebibliography}{25}
\expandafter\ifx\csname natexlab\endcsname\relax\def\natexlab#1{#1}\fi
\expandafter\ifx\csname bibnamefont\endcsname\relax
  \def\bibnamefont#1{#1}\fi
\expandafter\ifx\csname bibfnamefont\endcsname\relax
  \def\bibfnamefont#1{#1}\fi
\expandafter\ifx\csname citenamefont\endcsname\relax
  \def\citenamefont#1{#1}\fi
\expandafter\ifx\csname url\endcsname\relax
  \def\url#1{\texttt{#1}}\fi
\expandafter\ifx\csname urlprefix\endcsname\relax\def\urlprefix{URL }\fi
\providecommand{\bibinfo}[2]{#2}
\providecommand{\eprint}[2][]{\url{#2}}

\bibitem[{\citenamefont{Onsager}(1931)}]{Onsager-31}
\bibinfo{author}{\bibfnamefont{L.}~\bibnamefont{Onsager}},
  \bibinfo{journal}{Phys. Rev.} \textbf{\bibinfo{volume}{38}},
  \bibinfo{pages}{2265} (\bibinfo{year}{1931}).

\bibitem[{\citenamefont{Ford and O'Connell}(1996)}]{Ford-96}
\bibinfo{author}{\bibfnamefont{G.~W.} \bibnamefont{Ford}} \bibnamefont{and}
  \bibinfo{author}{\bibfnamefont{R.~F.} \bibnamefont{O'Connell}},
  \bibinfo{journal}{Phys. Rev. Lett.} \textbf{\bibinfo{volume}{77}},
  \bibinfo{pages}{798} (\bibinfo{year}{1996}).

\bibitem[{\citenamefont{{Lax}}(2000)}]{Lax-00}
\bibinfo{author}{\bibfnamefont{M.}~\bibnamefont{{Lax}}},
  \bibinfo{journal}{Optics Communications} \textbf{\bibinfo{volume}{179}},
  \bibinfo{pages}{463} (\bibinfo{year}{2000}).

\bibitem[{\citenamefont{Callen and Welton}(1951)}]{Callen-51}
\bibinfo{author}{\bibfnamefont{H.~B.} \bibnamefont{Callen}} \bibnamefont{and}
  \bibinfo{author}{\bibfnamefont{T.~A.} \bibnamefont{Welton}},
  \bibinfo{journal}{Phys. Rev.} \textbf{\bibinfo{volume}{83}},
  \bibinfo{pages}{34} (\bibinfo{year}{1951}).

\bibitem[{\citenamefont{Abragam}(1961)}]{Abragam-61}
\bibinfo{author}{\bibfnamefont{A.}~\bibnamefont{Abragam}},
  \emph{\bibinfo{title}{Principles of Nuclear Magnetism}}
  (\bibinfo{publisher}{Oxford University Press}, \bibinfo{address}{Oxford},
  \bibinfo{year}{1961}).

\bibitem[{\citenamefont{Fine}(1997)}]{Fine-97}
\bibinfo{author}{\bibfnamefont{B.~V.} \bibnamefont{Fine}},
  \bibinfo{journal}{Phys. Rev. Lett.} \textbf{\bibinfo{volume}{79}},
  \bibinfo{pages}{4673} (\bibinfo{year}{1997}).

\bibitem[{\citenamefont{Zhang et~al.}(2007)\citenamefont{Zhang, Konstantinidis,
  {Al-Hassanieh}, and Dobrovitski}}]{Zhang-07}
\bibinfo{author}{\bibfnamefont{W.}~\bibnamefont{Zhang}},
  \bibinfo{author}{\bibfnamefont{N.}~\bibnamefont{Konstantinidis}},
  \bibinfo{author}{\bibfnamefont{K.~A.} \bibnamefont{{Al-Hassanieh}}},
  \bibnamefont{and} \bibinfo{author}{\bibfnamefont{V.~V.}
  \bibnamefont{Dobrovitski}}, \bibinfo{journal}{J. Phys. Condens. Matter}
  \textbf{\bibinfo{volume}{19}}, \bibinfo{pages}{083202}
  (\bibinfo{year}{2007}).

\bibitem[{\citenamefont{Verstraete and Cirac}(2004)}]{Verstraete-04}
\bibinfo{author}{\bibfnamefont{F.}~\bibnamefont{Verstraete}} \bibnamefont{and}
  \bibinfo{author}{\bibfnamefont{J.}~\bibnamefont{Cirac}},
  \bibinfo{journal}{cond-mat/0407066}  (\bibinfo{year}{2004}).

\bibitem[{\citenamefont{White and Feiguin}(2004)}]{White-04}
\bibinfo{author}{\bibfnamefont{S.~R.} \bibnamefont{White}} \bibnamefont{and}
  \bibinfo{author}{\bibfnamefont{A.~E.} \bibnamefont{Feiguin}},
  \bibinfo{journal}{Phys. Rev. Lett.} \textbf{\bibinfo{volume}{93}},
  \bibinfo{pages}{076401} (\bibinfo{year}{2004}).

\bibitem[{\citenamefont{Schollw{\"o}ck}(2005)}]{Schollwock-05}
\bibinfo{author}{\bibfnamefont{U.}~\bibnamefont{Schollw{\"o}ck}},
  \bibinfo{journal}{Rev. Mod. Phys.} \textbf{\bibinfo{volume}{77}},
  \bibinfo{pages}{259} (\bibinfo{year}{2005}).

\bibitem[{\citenamefont{Crosswhite et~al.}(2008)\citenamefont{Crosswhite,
  Doherty, and Vidal}}]{Vidal-08}
\bibinfo{author}{\bibfnamefont{G.~M.} \bibnamefont{Crosswhite}},
  \bibinfo{author}{\bibfnamefont{A.~C.} \bibnamefont{Doherty}},
  \bibnamefont{and} \bibinfo{author}{\bibfnamefont{G.}~\bibnamefont{Vidal}},
  \bibinfo{journal}{Phys. Rev. B} \textbf{\bibinfo{volume}{78}},
  \bibinfo{pages}{035116} (\bibinfo{year}{2008}).

\bibitem[{\citenamefont{Gemmer et~al.}(2004)\citenamefont{Gemmer, Michel, and
  Mahler}}]{Gemmer-04}
\bibinfo{author}{\bibfnamefont{J.}~\bibnamefont{Gemmer}},
  \bibinfo{author}{\bibfnamefont{M.}~\bibnamefont{Michel}}, \bibnamefont{and}
  \bibinfo{author}{\bibfnamefont{G.}~\bibnamefont{Mahler}},
  \emph{\bibinfo{title}{Quantum Thermodynamics}} (\bibinfo{publisher}{Springer
  Verlag}, \bibinfo{address}{Berlin/Heidelberg}, \bibinfo{year}{2004}).

\bibitem[{\citenamefont{Goldstein et~al.}(2006)\citenamefont{Goldstein,
  Leibowitz, Tumulka, and Zanghi}}]{Goldstein-06}
\bibinfo{author}{\bibfnamefont{S.}~\bibnamefont{Goldstein}},
  \bibinfo{author}{\bibfnamefont{J.~L.} \bibnamefont{Leibowitz}},
  \bibinfo{author}{\bibfnamefont{R.}~\bibnamefont{Tumulka}}, \bibnamefont{and}
  \bibinfo{author}{\bibfnamefont{N.}~\bibnamefont{Zanghi}},
  \bibinfo{journal}{Phys. Rev. Lett.} \textbf{\bibinfo{volume}{96}},
  \bibinfo{pages}{050403} (\bibinfo{year}{2006}).

\bibitem[{\citenamefont{Popescu et~al.}(2006)\citenamefont{Popescu, Short, and
  Winter}}]{Popescu-06}
\bibinfo{author}{\bibfnamefont{S.}~\bibnamefont{Popescu}},
  \bibinfo{author}{\bibfnamefont{A.~J.} \bibnamefont{Short}}, \bibnamefont{and}
  \bibinfo{author}{\bibfnamefont{A.}~\bibnamefont{Winter}},
  \bibinfo{journal}{Nature Physics} \textbf{\bibinfo{volume}{2}},
  \bibinfo{pages}{754} (\bibinfo{year}{2006}).

\bibitem[{\citenamefont{Reimann}(2007)}]{Reimann-07}
\bibinfo{author}{\bibfnamefont{P.}~\bibnamefont{Reimann}},
  \bibinfo{journal}{Phys. Rev. Lett.} \textbf{\bibinfo{volume}{99}},
  \bibinfo{pages}{160404} (\bibinfo{year}{2007}).

\bibitem[{\citenamefont{\'Alvarez et~al.}(2008)\citenamefont{\'Alvarez,
  Danieli, Levstein, and Pastawski}}]{Alvarez-08}
\bibinfo{author}{\bibfnamefont{G.~A.} \bibnamefont{\'Alvarez}},
  \bibinfo{author}{\bibfnamefont{E.~P.} \bibnamefont{Danieli}},
  \bibinfo{author}{\bibfnamefont{P.~R.} \bibnamefont{Levstein}},
  \bibnamefont{and} \bibinfo{author}{\bibfnamefont{H.~M.}
  \bibnamefont{Pastawski}}, \bibinfo{journal}{Phys. Rev. Lett.}
  \textbf{\bibinfo{volume}{101}}, \bibinfo{pages}{120503}
  (\bibinfo{year}{2008}).

\bibitem[{\citenamefont{Bartsch and Gemmer}(2009)}]{Bartsch-09}
\bibinfo{author}{\bibfnamefont{C.}~\bibnamefont{Bartsch}} \bibnamefont{and}
  \bibinfo{author}{\bibfnamefont{J.}~\bibnamefont{Gemmer}},
  \bibinfo{journal}{Phys. Rev. Lett.} \textbf{\bibinfo{volume}{102}},
  \bibinfo{pages}{110403} (\bibinfo{year}{2009}).

\bibitem[{\citenamefont{Bartsch and Gemmer}(2011)}]{Bartsch-11}
\bibinfo{author}{\bibfnamefont{C.}~\bibnamefont{Bartsch}} \bibnamefont{and}
  \bibinfo{author}{\bibfnamefont{J.}~\bibnamefont{Gemmer}},
  \bibinfo{journal}{Europhys. Lett.} \textbf{\bibinfo{volume}{96}},
  \bibinfo{pages}{60008} (\bibinfo{year}{2011}).

\bibitem[{\citenamefont{Long et~al.}(2003)\citenamefont{Long, Prelov{\v{s}}ek,
  {El Shawish}, Karadamoglou, and Zotos}}]{Long-03}
\bibinfo{author}{\bibfnamefont{M.~W.} \bibnamefont{Long}},
  \bibinfo{author}{\bibfnamefont{P.}~\bibnamefont{Prelov{\v{s}}ek}},
  \bibinfo{author}{\bibfnamefont{S.}~\bibnamefont{{El Shawish}}},
  \bibinfo{author}{\bibfnamefont{J.}~\bibnamefont{Karadamoglou}},
  \bibnamefont{and} \bibinfo{author}{\bibfnamefont{X.}~\bibnamefont{Zotos}},
  \bibinfo{journal}{Phys. Rev. B} \textbf{\bibinfo{volume}{68}},
  \bibinfo{pages}{235106} (\bibinfo{year}{2003}).

\bibitem[{\citenamefont{White}(2009)}]{White-09}
\bibinfo{author}{\bibfnamefont{S.~R.} \bibnamefont{White}},
  \bibinfo{journal}{Phys. Rev. Lett.} \textbf{\bibinfo{volume}{102}},
  \bibinfo{pages}{190601} (\bibinfo{year}{2009}).

\bibitem[{\citenamefont{Fine}(2009)}]{Fine-09}
\bibinfo{author}{\bibfnamefont{B.~V.} \bibnamefont{Fine}},
  \bibinfo{journal}{Phys. Rev. E} \textbf{\bibinfo{volume}{80}},
  \bibinfo{pages}{051130} (\bibinfo{year}{2009}).

\bibitem[{Nor()}]{Normalization}
\bibinfo{note}{Here we neglect the normalization correction $|\Peq|^2 = 1 + O
  (1/\sqrt{N})$ associated with the statistical fluctuations of $|a_k|^2$.}

\bibitem[{Sup()}]{Supplement}
\bibinfo{note}{See the supplemental material.}

\bibitem[{\citenamefont{Bloch}(1946)}]{Bloch-46}
\bibinfo{author}{\bibfnamefont{F.}~\bibnamefont{Bloch}},
  \bibinfo{journal}{Phys. Rev.} \textbf{\bibinfo{volume}{70}},
  \bibinfo{pages}{460} (\bibinfo{year}{1946}).

\bibitem[{\citenamefont{de~Alcantara~Bonfim and Reiter}(1992)}]{Bonfim-92}
\bibinfo{author}{\bibfnamefont{O.~F.} \bibnamefont{de~Alcantara~Bonfim}}
  \bibnamefont{and} \bibinfo{author}{\bibfnamefont{G.}~\bibnamefont{Reiter}},
  \bibinfo{journal}{Phys. Rev. Lett.} \textbf{\bibinfo{volume}{69}},
  \bibinfo{pages}{367} (\bibinfo{year}{1992}).

\end{thebibliography}

\ 

\

\centerline{\bf SUPPLEMENTAL MATERIAL}

\

\setcounter{figure}{0}
\renewcommand{\thefigure}{S\arabic{figure}}

\setcounter{equation}{0}
\renewcommand{\theequation}{S\arabic{equation}}

Note: Equation numbers, figure numbers and citation numbers appearing in this supplemental material start with letter ``S". Equation, figure and citation numbers without ``S" refer to the text of the main article. 

\section{Derivation of Eqs.(6,7)}

To obtain Eq.(6), the correlation function (5)  can be represented in the energy eigenbasis as follows:
\begin{equation}
C(t,T) = {1 \over 2T} \int_{-T}^{T} dt' \sum_{m,n,p,q} a_q^* a_n^* a_p a_m  A_{qp} A_{nm} e^{i\left(\epsilon_q-\epsilon_p \right )t} e^{i\left(\epsilon_q-\epsilon_p +\epsilon_n-\epsilon_m \right )t'}
\label{fluct2-matr}
\end{equation}
 Due to the random phases of complex amplitudes $a_n$, the Hilbert-space average $\overline{a_q^* a_n^* a_p a_m }$ is not zero only when the four amplitudes form conjugate pairs such as $\overline{a_q^* a_q a_n^* a_n }$. The main contribution to the RHS of Eq.(\ref{fluct2-matr}) comes from the 2-pair terms where the indices $q$ and $n$ are different from each other, in which case, according to Eq.(4), $\overline{a_q^* a_q a_n^* a_n  } = (\overline{|a_q|^2})^2 = 1/N^2 $. The number of the remaining terms, where the two pair indices coincide, is smaller by factor $1/N$, and therefore, their contribution to the RHS of Eq.(\ref{fluct2-matr}) is suppressed by the same factor as long as the set of the eigenvalues of  $\Ahat$  does not have a small subset of the anomalously large members. Limiting ourselves to the terms with different pair indices, we obtain
\begin{equation}
\overline{ a_q^* a_n^* a_p a_m} = 1/N^2 (\delta_{qp} \delta_{nm} + \delta_{qm} \delta_{np})
\label{4a}
\end{equation}
The term $\delta_{qp} \delta_{nm}$ in the parentheses does not make contribution to the average because of the condition $\Tr\{\Ahat\} = 0$. The summation over the term $ \delta_{qm} \delta_{np} $ with the subsequent integration over $t'$ gives 
\begin{equation}
\overline{C(t,T)} = {1 \over N^2} \sum_{m,n}e^{i\left ( \epsilon_m-\epsilon_n \right )t}A_{mn}A_{nm} = {1 \over N^2} \Tr \left\{ \Ahat(t) \Ahat(0) \right\},
\label{Cav}
\end{equation} 
which is the principal term in the RHS of Eq.(6). 

The deviation of $C(t,T)$ obtained for a particular realization of $\Peq$ from $\overline{C(t, T)}$ can be estimated from the variance $\overline{\Delta_1^2} \equiv \overline{\left[ C(t, T) - \overline{C(t, T)} \right]^2}$. The explicit form of this variance obtained using Eq.(\ref{4a}) is
\begin{eqnarray}
& \overline{\Delta_1^2} = {1 \over 4T^2} \int_{-T}^{T} dt' \int_{-T}^{T} dt'' 
A_{qp} A_{nm} A_{kl} A_{uv} \ e^{i\left(\epsilon_q-\epsilon_p +\epsilon_k - \epsilon_l \right )t} \  
e^{i\left(\epsilon_q-\epsilon_p +\epsilon_n-\epsilon_m \right )t'} \ 
e^{i\left(\epsilon_k-\epsilon_l +\epsilon_u-\epsilon_v \right )t''}
&
\nonumber
\\
&
\times
\sum_{q,p,m,n,k,l,u,v} \left[ \overline{a_q^* a_n^* a_k^* a_u^* a_p a_m a_l a_v }
-  {1\over N^4} (\delta_{qp} \delta_{nm} + \delta_{qm} \delta_{np})(\delta_{kl} \delta_{uv} + \delta_{kv} \delta_{ul})\right] .
&
\label{fluct2-var}
\end{eqnarray}
The evaluation of $\overline{a_q^* a_n^* a_k^* a_u^* a_p a_m a_l a_v }$ relies on the same considerations as those that led to Eq.(\ref{4a}). Namely, the nonzero contribution to the sum in the RHS of Eq.(\ref{fluct2-var}) comes from the terms, where the eight amplitudes are  organized into conjugate pairs such as $\overline{a_q^* a_q a_n^* a_n  a_k^* a_k a_u^* a_u }$. The main contribution to the RHS of Eq.(\ref{fluct2-var}) comes from the 4-pair terms which have all four indices $q,n,k$ and $l$ different from each other.  In this case, according to Eq.(4), $\overline{a_q^* a_q a_n^* a_n  a_k^* a_k a_u^* a_u } = (\overline{|a_q|^2})^4 = 1/N^4 $. Now we observe that the number of all possible combinations of four different conjugate pairs in expression $\overline{a_q^* a_n^* a_k^* a_u^* a_p a_m a_l a_v }$ is relatively large. However, we find by inspection that most of these combinations eventually give zero contribution to the RHS of Eq.(\ref{fluct2-var}) because of the condition $\Tr\{\Ahat\} = 0$. Below, we only include those combinations that give nonzero contributions:
\begin{eqnarray}
\overline{a_q^* a_n^* a_k^* a_u^* a_p a_m a_l a_v } &= {1\over N^4} [ & 
 \delta_{qm} \delta_{np} \delta_{kv} \delta_{ul} 
\ 
+ \  \delta_{ql} \delta_{nv} \delta_{kp} \delta_{um}
\
+ \ \delta_{qv} \delta_{nl} \delta_{km} \delta_{up}
\label{8a-average}
\\
\nonumber
&&
+ \  \delta_{qm} \delta_{nl} \delta_{kv} \delta_{up} 
\ 
+ \  \delta_{qv} \delta_{np} \delta_{km} \delta_{ul}
\
+ \  \delta_{qm} \delta_{nv} \delta_{kp} \delta_{ul}
\\
\nonumber
&&  
+ \  \delta_{ql} \delta_{np} \delta_{kv} \delta_{um}
\
+ \  \delta_{ql} \delta_{nv} \delta_{km} \delta_{up}
\
+ \  \delta_{qv} \delta_{nl} \delta_{kp} \delta_{um}
\\
\nonumber
&& 
+ \hbox{ \{omitted terms\} }]
.
\end{eqnarray}

\begin{figure} \setlength{\unitlength}{0.1cm}

\begin{picture}(100, 45)
{ 
\put(0,38){\textsf{ (a)}}
\put(0, 0){ \epsfxsize= 1.5in \epsfbox{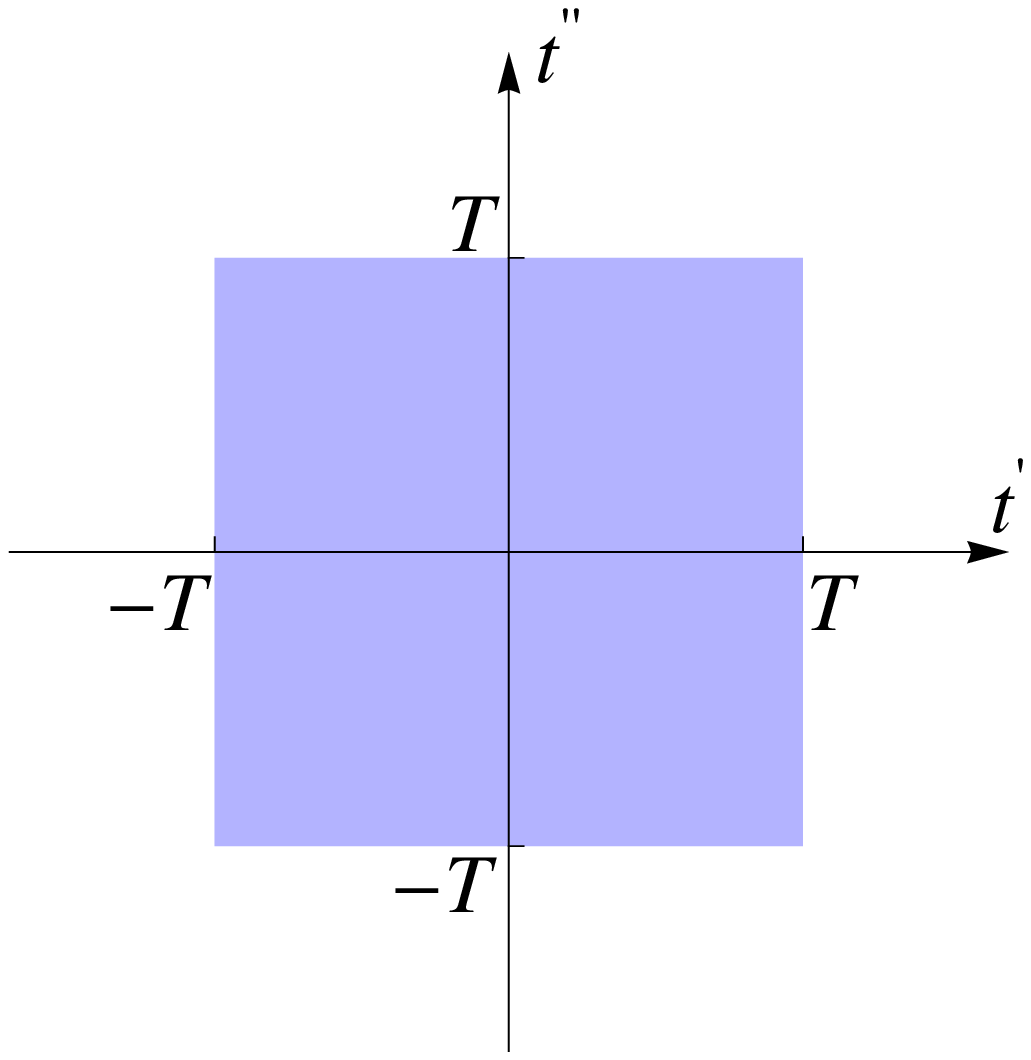} }
\put(40,38){\textsf{(b)}}
\put(40, 0){ \epsfxsize= 1.5in \epsfbox{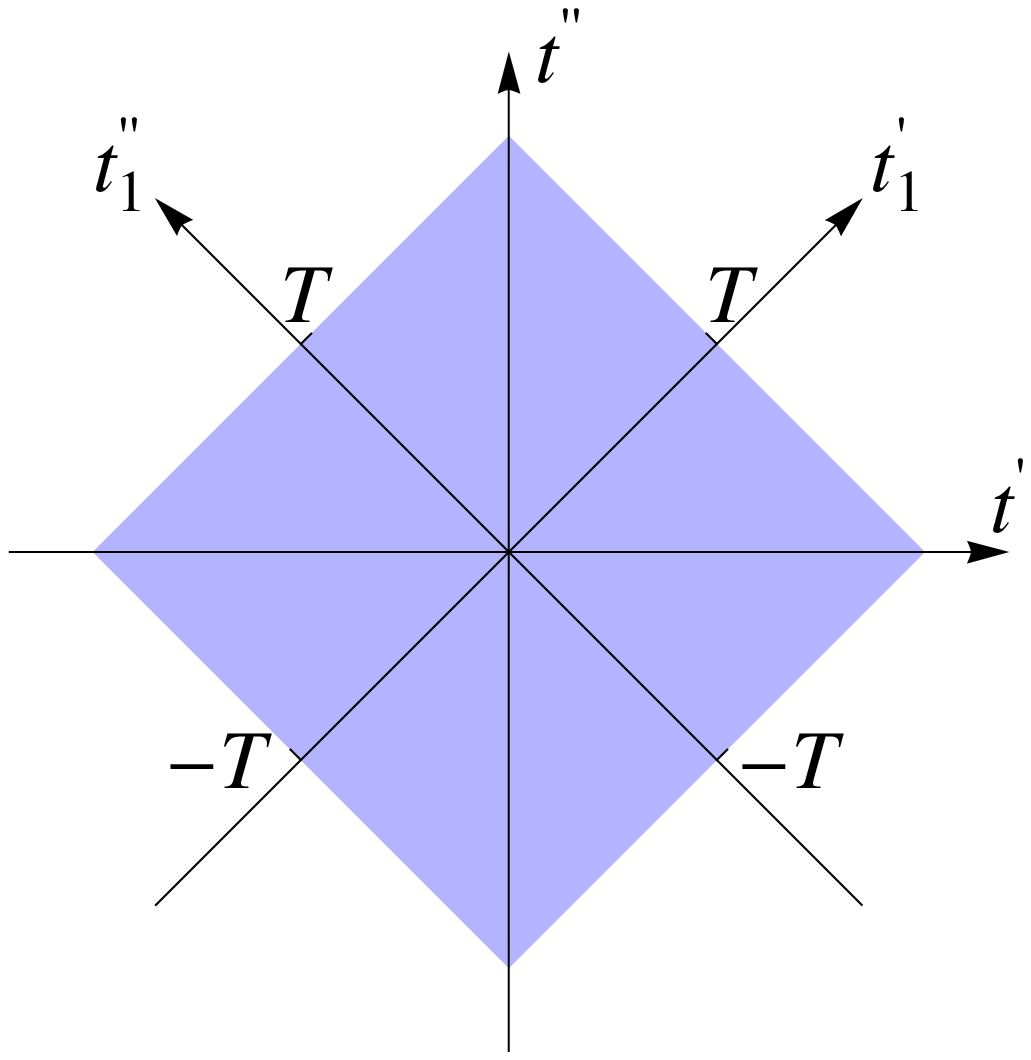} }
}
\end{picture} 
\caption{(Color online)Figures illustrating the rotation of coordinates and the rotation of the integration region. (a) Original coordinates $(t',t'')$ and the original integration region. (b) Rotated coordinates$(t^{'}_1, t^{''}_2)$ and rotated integration region. 
} 
\label{fig-region} 
\end{figure}

After substituting Eq.(\ref{8a-average}) into Eq.(\ref{fluct2-var}) most of the resulting terms contain integral of the type:
\begin{equation}
{1 \over 4T^2} \int_{-T}^{T} \int_{-T}^{T} dt' dt'' 
\  
e^{i \eta t'} \ 
e^{-i \eta t''},
\label{region1}
\end{equation}
where $\eta$ is some combination of energies $\epsilon_n$. The integration region for the above integral is shown in Fig.\ref{fig-region}(a). Since our primary goal is to find the order of magnitude of the leading terms in the limit of large $T$, we adopt the following relatively crude approximation, which significantly simplifies the analytic expressions. Namely, we rotate the integration region as shown in Fig.\ref{fig-region}(b) and simultaneously change the integration variables to $t^{'}_1 = (t' + t'')/\sqrt{2}$ and $t^{''}_1 = (t'' - t')/\sqrt{2}$. Finally, we substitute $t_2 = t^{''}_1 \sqrt{2}$, thereby replacing  expression (\ref{region1}) with
\begin{equation}
{1 \over 2T \sqrt{2}} \int_{-T\sqrt{2}}^{T\sqrt{2}} dt_2 
\  
e^{-i \eta t_2}.
\label{region2}
\end{equation} 

Among the nine terms in the square brackets in Eq.(\ref{8a-average}), not all terms contribute equally to the RHS of Eq.(\ref{fluct2-var}). The main contribution comes from the first, the second and the third terms. The contribution from the first term is then canceled exactly by  the contribution of the term containing $1/N^4$ in the square brackets in Eq.(\ref{fluct2-var}). What remains after that is:
\begin{equation}
 \overline{\Delta_1^2} \approx 
{1 \over 2 \sqrt{2}\ T N^4} \int_{-T\sqrt{2}}^{T\sqrt{2}} dt_2 
\left( 
\left[ \Tr \left\{ \Ahat(t_2) \Ahat(0) \right\} \right]^2
+ \Tr \left\{ \Ahat (t- t_2) \Ahat(0) \right\} \Tr \left\{ \Ahat (t+ t_2) \Ahat(0) \right\}
\right).
\label{fluct2-var-result}
\end{equation}
The two terms in the above integral originate, in their respective order, from the second and the third terms in Eq.(\ref{8a-average}). Each of the remaining six terms in Eq.(\ref{8a-average}) would contribute to the RHS of Eq.(\ref{fluct2-var}) a term of the following type:  $\Tr \left\{ \Ahat(t) \Ahat (0) \Ahat(t+ t_2) \Ahat(t_2) \right\}$. For $t_2 \sim T \gg t$, the operators separated by $t_2$ should become uncorrelated, which would imply that the above term becomes equal to ${1 \over N} \left[ \Tr \left\{ \Ahat(t) \Ahat(0) \right\} \right]^2$, which is smaller than the terms included in the estimate (\ref{fluct2-var-result}) by factor $1/N$.

In general, the two terms in Eq.(\ref{fluct2-var-result}) are of the same order of magnitude, when $t$ is of the order of the characteristic decay time of the correlation function $\Tr \left\{ \Ahat (t) \Ahat(0) \right\}$. However, if $\Tr \left\{ \Ahat (t) \Ahat(0) \right\}$ decays asymptotically faster than $|t|^{-0.5}$, then  the second term in Eq.(\ref{fluct2-var-result}) becomes much smaller than the first one as $t$ increases.

Since the above derivation is done in the leading order in $1/N$, it implicitly assumes that $T$ is much smaller than the inverse level spacing of the system. If one is to consider the limit $T \to \infty$ at fixed $N$, then the contributions from the corrections of higher order in $1/N$ would need to be examined. 


\section{Derivation of Eqs.(9,10)}

The derivation of Eq.(9) is based on the following general relationships: If $\Vhat$ is a Hermitian operator, then we can use the eigenbasis of $\Vhat$ to define $|\Peq \rangle$ in Eq.(3) and then also use the probability distribution (4) to obtain  $\overline{ \langle \Peq | \Vhat|  \Peq \rangle } = \sum_{m}\overline{|a_m|^2} V_{mm} =  {1 \over N} \Tr \left\{ \Vhat \right\}$.  The variance with respect to the above average is:
$\overline{\Delta_2^2} = \sum_{m}\overline{(|a_m|^2 - 1/N)^2} \ V_{mm}^2 = {1\over N^2} \Tr \left\{ \Vhat^2 \right\} $. (This formula is different from the closely related Eq.(2) of Ref.[17], because, as noted in [22],  we allow statistical noise for the normalization of $|  \Peq \rangle$.) In the context of Eq.(9), we can choose $\Vhat = \Ahat (t) \Ahat(0)$ and thereby obtain Eq.(10).

\section{Comparison with the relaxation of classical spins}

In order to illustrate the computational advantage of the quantum parallelism associated with the direct evaluation of $\langle \Pneq | \Ahat (t)|\Pneq \rangle $, let us consider what it takes to  calculate the high-temperature relaxation function of the $z$-component of the total magnetization  of classical spin systems  numerically.

Let us assume that the system investigated numerically is a chain of $N_s$ classical spins governed by the Hamiltonian 
\begin{equation}
\Hm=\sum_{i<j}^{N_s} J_{ij}^x S_{ix} S_{jx}+ J_{ij}^y S_{iy} S_{jy}+J_{ij}^z S_{iz} S_{jz}~,
\label{H}
\end{equation} 
where $(S_{ix}, S_{iy}, S_{iz})$ are the projections of classical spin vectors having length  $\sqrt{S_{ix}^2 + S_{iy}^2 + S_{iz}^2} =1 $, and $J_{ij}^x , J_{ij}^y , J_{ij}^z$ are the anisotropic coupling constants. 

We are interested in the linear response regime. Hence the initial polarization of the spin system should be much smaller than $N_s$. Let us take it to be equal to $0.1 N_s$. At the same time, the rms magnetization noise level for this system at the infinite temperature equilibrium is $\sqrt{N_s\left\langle {S_m^z}^2 \right\rangle}=\sqrt{N_s/3}$. Thus, in a single numerical realization of the relaxation process, one can only obtain a reasonable accuracy in the range between $0.1 N_s$ and roughly $2 \sqrt{N_s/3}$. For 1000 classical spins, this means that a single run of a relaxation process will only be statistically accurate in the range between 100 and 40.  The standard way to improve the quality of the calculation is to average over $n_0$ independent realizations of the relaxation process. The statistical noise in this case decreases very slowly --- only as $1/\sqrt{n_0}$. 

If one computes $\langle \Pneq | \Ahat (t)|\Pneq \rangle $ for a single realization of $|\Pneq \rangle$ for $N_s$ spins 1/2 with the initial 10 percent polarization, i.e. $\langle \Pneq | \Ahat (0)|\Pneq \rangle = 0.1 \times {1\over 2} N_s $, then this single run would only be affected by the statistical noise associated with $\langle \Peq | \Ahat (t)|\Peq \rangle $ when $\langle \Pneq | \Ahat (t)|\Pneq \rangle$ decays to the values of the order of ${\sqrt{N_s\left\langle {S_m^z}^2 \right\rangle} \over 2^{N_s/2}}$. For $N_s= 1000$, such a noise would be too small to be of any practical concern. However, for a system of 20 spins, the equilibrium noise does have observable effect on  $\langle \Pneq | \Ahat (t)|\Pneq \rangle $, as can be seen in Fig.~1. As in the classical case, this noise can be further suppressed  by factor $1/\sqrt{n_0}$ after averaging over $n_0$ independent realizations of $\langle \Pneq | \Ahat (t)|\Pneq \rangle$.

\section{Monitoring $\Ahat$}

The fluctuations of $\langle \Peq | \Ahat (t)|\Peq \rangle $ should be distinguished from the measured equilibrium noise, when a macroscopic quantum observable $\Ahat$ is continuously monitored[S1-S4]. Let us again assume that $\Ahat$ represents the total $z$-component of the magnetization of the spin system. In this case, the rms amplitude of the {\it monitored} infinite-temperature spin noise is expected to be $\sqrt{{1 \over N} \Tr \{ \Ahat^2 \}} = \sqrt{N_s\left\langle {S_m^z}^2 \right\rangle}$ [24], while, as implied by Eq.(6), the rms amplitude of the quantum mechanical expectation value is suppressed by the extra factor $1/\sqrt{N}$. 
As mentioned in the main article, this suppression is the consequence of the fact that
the noise of the quantum-mechanical expectation value is, in a sense, averaged over the quantum superposition of $N$ independent ``noises'' associated with the time evolution of each of the basis states in the Schroedinger representation. 

Despite the exponentially large difference in the amplitudes,  the monitored noise and the noise of $\langle \Peq | \Ahat (t)|\Peq \rangle $ should have identical normalized time correlation functions in {\it macroscopic} systems.  If the total magnetization is the only variable being continuously monitored in the system (or one of a small number of variables), then the disturbance of the dynamics of an individual spin in the bulk of the sample should be minimal. Since the fluctuations of the total magnetization come mostly from spins in the bulk, this means that  the correlation function of the monitored noise should still be proportional to the quantum mechanical trace $\Tr \left\{ \Ahat(t) \Ahat(0) \right\}$ computed for a perfectly isolated quantum system. 

{\small

\

[S1] W. M. Itano {\it  et al.}, Phys. Rev. A {\bf 47}, 3554-3570 (1993).

[S2] J. L. Sorensen, J. Hald  and E. S. Polzik, Phys. Rev. Lett. {\bf 80}, 3487 (1998).

[S3] T. Sleator, E. L. Hahn, C. Hilbert, and J. Clarke, Phys. Rev. Lett. {\bf 55}, 1742 (1985).

[S4] S. A. Crooker, D. G. Rickel, A. V. Balatsky and D. L. Smith, Nature {\bf 431}, 49 (2004).
}

\end{document}